\documentclass[prc,twocolumn,floatfix,groupedaddress,nofootinbib,showpacs,preprintnumbers,
amsmath,amssymb,amsfonts,widetable] {revtex4}
\usepackage{bm}
\usepackage{mathrsfs}
\usepackage{amssymb}
\usepackage{amsmath}
\usepackage{graphicx}
\usepackage{array}
\usepackage{lipsum}

\def\alphad{$\!\alpha_{\raisebox{-1pt}{\tiny  D}}\,$}

\begin{document}
%%%%%%%%%%%
%%%%%%%%%%%
\title{Difference in proton radii of mirror nuclei\\
as a possible surrogate for the neutron skin}
%%%
\author{Junjie Yang}%\email{jy14f@my.fsu.edu}
\author{J. Piekarewicz}%\email{jpiekarewicz@fsu.edu}
\affiliation{Department of Physics, Florida State University,
               Tallahassee, FL 32306, USA}
%%%%%%%%%%%
\date{\today}
\begin{abstract}
 \noindent
  It has been recently suggested that differences in the charge radii of mirror nuclei are 
  proportional to the neutron-skin thickness of neutron-rich nuclei and to the slope of the 
  symmetry energy $L$ [B.A. Brown Phys. Rev. Lett. 119, 122502 (2017)]. The determination
  of the neutron skin has important implications for nuclear physics and astrophysics. Although 
  the use of electroweak probes provides a largely model-independent determination of 
  the neutron skin, the experimental challenges are enormous. Thus, the possibility
  that differences in the charge radii of mirror nuclei may be used as a surrogate for 
  the neutron skin is a welcome alternative. To test the validity of this assumption we 
  perform calculations based on a set of relativistic energy density functionals that 
  span a wide region of values of $L$. Our results confirm that the difference in charge 
  radii of various neutron-deficient nickel isotopes and their corresponding mirror nuclei 
  is indeed strongly correlated to both the neutron-skin thickness and $L$. Moreover, 
  given that various neutron-star properties are also sensitive to $L$, a data-to-data
  relation emerges between the difference in charge radii of mirror nuclei and the
  radius of low-mass neutron stars. 
 \end{abstract}
\smallskip
\maketitle

The neutron-rich skin of medium and heavy nuclei is a fundamental nuclear
property that has gained prominence almost two decades ago because of 
its strong correlation to the equation of state of neutron-rich matter, primarily 
to the slope of the symmetry energy $L$\,\cite{Brown:2000, 
Furnstahl:2001un, Centelles:2008vu,RocaMaza:2011pm}. 
Given that the weak charge of the neutron is significantly larger than the 
corresponding one for the proton, parity-violating electron scattering 
offers a clean probe of neutron densities that is free from strong-interaction 
uncertainties\,\cite{Donnelly:1989qs}. The pioneering Lead Radius Experiment 
(PREX) at the Jefferson Laboratory has provided the first model-independent 
evidence on the existence of a neutron-rich skin in 
${}^{208}$Pb\,\cite{Abrahamyan:2012gp,Horowitz:2012tj}. In the near future
a follow-up experiment (PREX-II) is envisioned to reach the desired $0.06$\,fm 
sensitivity in the neutron radius of ${}^{208}$Pb and a brand new experiment 
on ${}^{48}$Ca promises to bridge the gap between modern ab initio approaches 
and density functional theory\,\cite{CREX:2013}. Moreover, the study of 
neutron-rich matter with unusual features such as large neutron skins is one 
of the key science drivers for the Facility for Rare Isotope 
Beams (FRIB)\,\cite{Balantekin:2014opa,LongRangePlan}.

Besides being of fundamental importance in nuclear structure, the neutron-rich 
skin of medium to heavy nuclei plays a critical role in the determination of the 
equation of state (EOS) of neutron-rich matter. In turn, important dynamical
signatures observed in the collision of heavy ions are encoded in the 
EOS\,\cite{Tsang:2004zz,Chen:2004si,Steiner:2005rd,
Shetty:2007zg,Tsang:2008fd,Li:2008gp,Tsang:2012se,Horowitz:2014bja}.
Further, despite a difference in length scales of 18 orders of magnitude, the 
neutron-skin thickness of ${}^{208}$Pb and the radius of a neutron star share 
a common dynamical origin\,\cite{Horowitz:2000xj,Horowitz:2001ya,
Carriere:2002bx,Steiner:2004fi,Li:2005sr,Erler:2012qd,Chen:2014sca}. Indeed,
the only input that the structure of spherically symmetric neutron stars is sensitive 
to is the equation of state of neutron-rich matter. This fact alone has created a unique 
synergy between nuclear physics and astrophysics. 

Although there is little doubt that parity violating electron scattering provides 
the cleanest probe of neutron densities, the experimental challenges associated 
with such experiments are enormous. This fact has motivated searches for 
complementary observables to the neutron skin that also display a strong 
sensitivity to the density dependence of the symmetry energy. Particularly 
valuable was the identification of the electric dipole polarizability (\,\alphad\!) 
as a strong isovector indicator\,\cite{Reinhard:2010wz}. The electric dipole 
polarizability encodes the response of the nucleus to an externally applied 
electric field and is directly proportional to the \emph{inverse} energy weighted 
sum of the isovector dipole response\,\cite{Harakeh:2001}. The isovector
dipole resonance is commonly identified as an out-of-phase oscillation of
protons against neutrons, with the symmetry energy acting as the restoring 
force. Since \alphad was first identified as a strong isovector indicator, a flurry 
of activity ensued in both theoretical\,\cite{Piekarewicz:2010fa,Piekarewicz:2012pp,
Reinhard:2012vw,Roca-Maza:2013mla,Roca-Maza:2015eza,Hagen:2015yea,
Piekarewicz:2006ip} and experimental fronts\,\cite{Tamii:2011pv,Poltoratska:2012nf,
Tamii:2013cna,Savran:2013bha,Hashimoto:2015ema,Rossi:2013xha,
Birkhan:2016qkr,Tonchev:2017ily}.

In the ongoing quest to determine the equation of state of neutron-rich matter,
B.A. Brown has recently identified a physical observable that is closely related 
to the neutron skin\,\cite{Brown:2017}. The argument is both simple and elegant: 
in the limit of exact charge symmetry, the neutron radius of a given nucleus is 
identical to the proton radius of its mirror nucleus. That is, 
%%%
\begin{align}
R_{\rm skin}(Z,N) \equiv& R_{n}(Z,N)-R_{p}(Z,N) \nonumber \\
\mathrel{\stackrel{\makebox[0pt]{\mbox{\normalfont\tiny c.s.}}}{=}} 
& R_{p}(N,Z)-R_{p}(Z,N)\!\equiv\!R_{\rm mirr}(Z,N).
\label{Mirror}
\end{align}
%%%
For example, in the case of ${}^{48}$Ca\,\cite{CREX:2013}:
%%%
\begin{align}
R_{\rm skin}({}^{48}{\rm Ca}) 
\mathrel{\stackrel{\makebox[0pt]{\mbox{\normalfont\tiny c.s.}}}{=}} 
R_{p}({}^{48}{\rm Ni})-R_{p}({}^{48}{\rm Ca})\equiv R_{\rm mirr}^{\,48}\,.
\label{Mirror48}
\end{align}
%%%

While the basic idea is appealing, the ultimate test of its validity relies on its 
robustness against the all-important Coulomb corrections. Indeed, most of 
the work in Ref.\,\cite{Brown:2017} was devoted to show that the differences 
in the charge radii of mirror nuclei as predicted by a set of Skyrme functionals 
is proportional to the slope of the symmetry energy at saturation density---even 
in the presence of Coulomb corrections. In this work we show that those 
findings remain valid in the relativistic approach. Moreover, we also show 
how $R_{\rm mirr}$, just as $R_{\rm skin}$, is correlated to the radius of 
low-mass neutron stars, a stellar property that is highly sensitive to 
the density dependence of the symmetry energy.

In the relativistic mean field (RMF) approach pioneered by Serot and
Walecka\,\cite{Walecka:1974qa,Serot:1984ey}, the basic fermionic
constituents are protons and neutrons interacting via photon exchange 
as well as through the exchange of various mesons of distinct Lorentz 
and isospin character. Besides the conventional Yukawa couplings of 
the mesons to the relevant nuclear currents, the model is supplemented 
by several nonlinear meson coupling that are essential for its ultimate
success\,\cite{Boguta:1977xi,Mueller:1996pm,Horowitz:2000xj}. 
Besides a progressive increase in the complexity of the model, sophisticated 
fitting protocols are now used for its calibration. Indeed, properties of finite 
nuclei, their monopole response, and even a few properties of neutron 
stars now provide critical inputs in the determination of the relativistic 
functional\,\cite{Chen:2014sca}. 

%%%%%%%%%%%%%%%%%%%%%%%%%%%%%%%%%%%%%%%%%%%%%%%%%%%%%
\begin{figure*}[ht]
\smallskip
 \includegraphics[width=0.65\columnwidth]{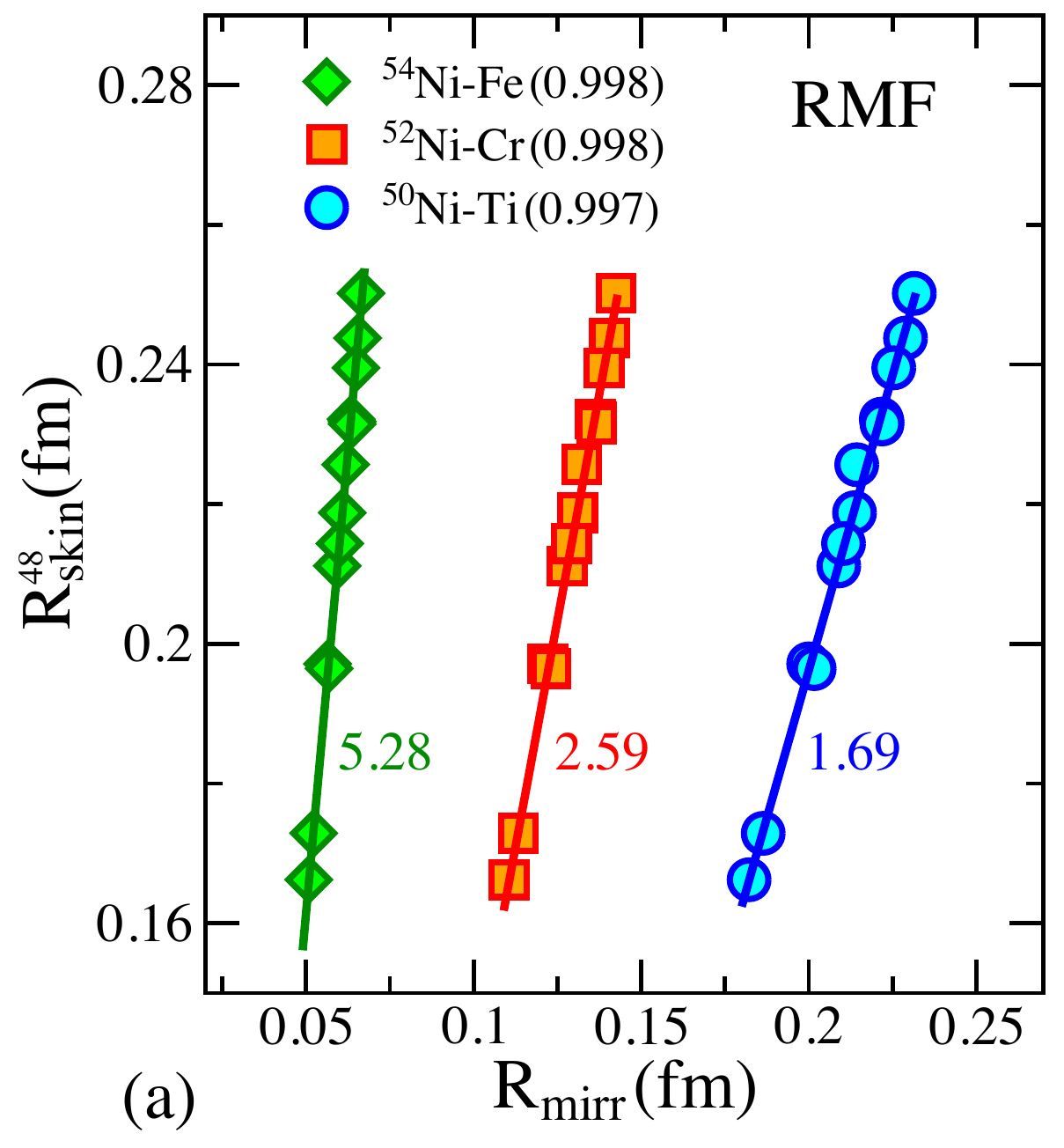}
   \hspace{10pt}
 \includegraphics[width=0.68\columnwidth]{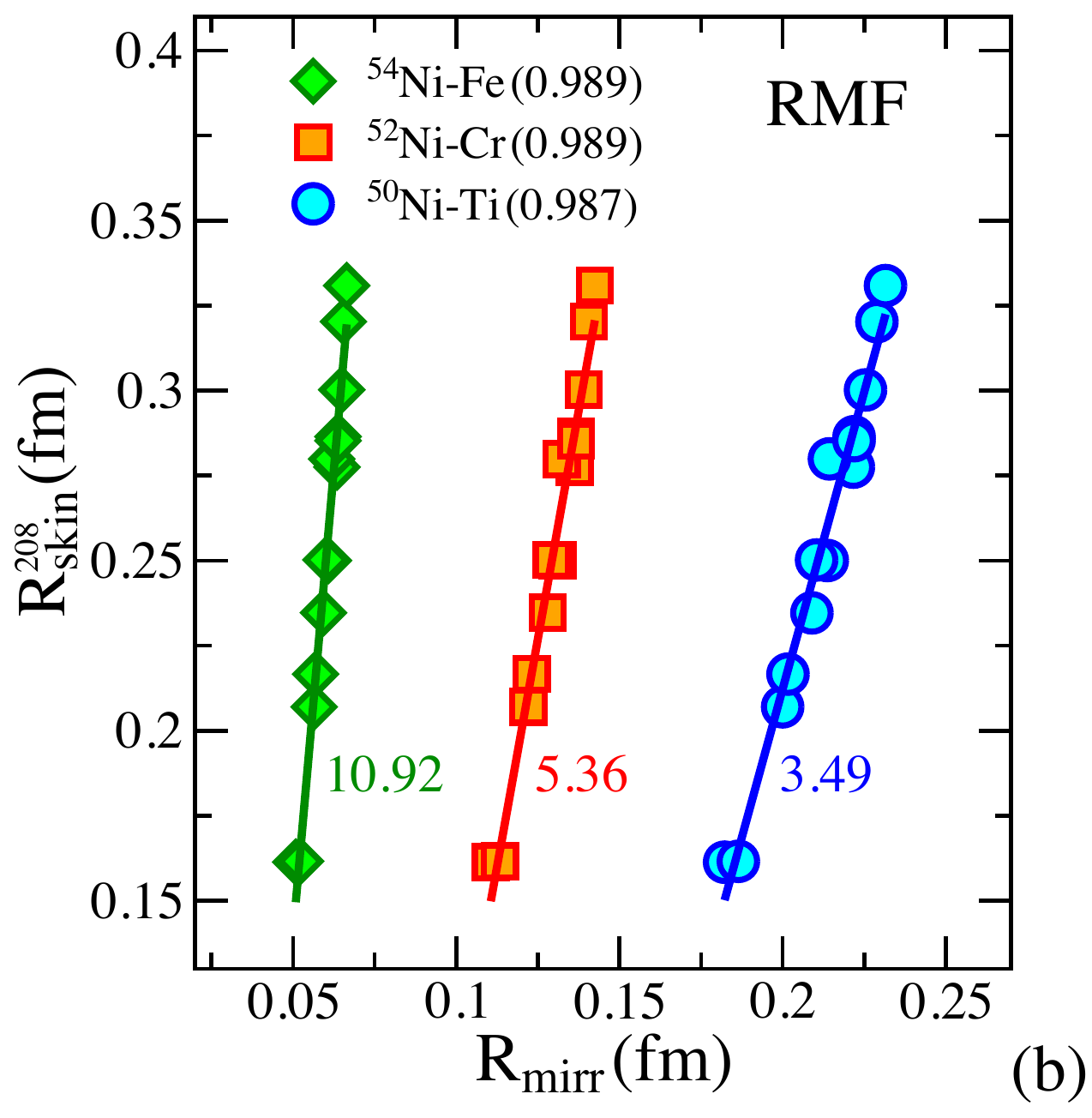}
\caption{(Color online) (a) Data-to-data relations between the neutron-skin thickness of
	       ${}^{48}$Ca and the difference in proton radii between a few neutron-deficient
	       nickel isotopes and their corresponding mirror nuclei along the 
	       $A\!=\!50, 52, 54$ isobars. Numbers in parentheses represent the
	       correlation coefficients and the numbers next  to the lines the linear regression 
	       slopes. (b) Same but now for the neutron-skin thickness of ${}^{208}$Pb.} 
\label{Fig1}
\end{figure*}
%%%%%%%%%%%%%%%%%%%%%%%%%%%%%%%%%%%%%%%%%%%%%%%%%%%%%

To explore some of the interesting correlations that emerged in Ref.\,\cite{Brown:2017}, 
but now in the relativistic context, we employ a set of 14 energy density functionals that 
span a wide region of values of the slope parameter: $L\!\simeq 50\!-\!140$\,MeV. In 
turn, this corresponds to a neutron-skin thickness in ${}^{208}$Pb ranging from about 
$R^{208}_{\rm skin}\!\simeq 0.15$ to $0.33$\,fm, well within the limits of the PREX 
measurement\,\cite{Abrahamyan:2012gp,Horowitz:2012tj}. Parameter sets for the 
models adopted in this contribution are: NL3\,\cite{Lalazissis:1996rd,Lalazissis:1999},
FSUGold\,\cite{Todd-Rutel:2005fa}, IU-FSU\,\cite{Fattoyev:2010mx}, 
TAMUC-FSU\,\cite{Fattoyev:2013yaa}, FSUGold2\,\cite{Chen:2014sca}, and
FSUGarnet\,\cite{Chen:2014mza}. Although most of these models have been 
accurately calibrated, a few of them were obtained by systematically varying 
their two isovector parameters, while leaving the isoscalar sector intact. This
enables one to modify the poorly constrained density dependence of the 
symmetry energy without compromising the success of the model in the
isoscalar sector. In essence, all these models reproduce nuclear observables 
near stability, yet may vary widely in their predictions for the properties of exotic 
nuclei far from stability.

In Fig.\,\ref{Fig1}(a) we display data-to-data relations between the neutron-skin
thickness of ${}^{48}$Ca and the difference in proton radii between three 
neutron-deficient nickel isotopes and their corresponding mirror nuclei; the same 
information is shown in Fig.\,\ref{Fig1}(b) but now for ${}^{208}$Pb. For example, 
the blue circles denote the difference in proton radii in the $A\!=\!50$ sector:
$R_{\rm mirr}^{50}\!\equiv\!R_{p}({}^{50}{\rm Ni})\!-\!R_{p}({}^{50}{\rm Ti})$.
Besides computing proton radii along the $A\!=\!50, 52, 54$ isobars, we tried
to calculate the proton radius of ${}^{48}$Ni but were unable to bind the $f^{7/2}$ 
protons---especially in RMF models having a soft symmetry energy. As indicated 
by the correlation coefficients displayed in parentheses in Fig.\,\ref{Fig1}, 
there is a strong correlation between $R_{\rm skin}$ and $R_{\rm mirr}$ for 
both ${}^{48}$Ca and ${}^{208}$Pb, at least for the representative set of models 
used in this work. Also shown are the linear regression slopes obtained from 
the statistical analysis. Clearly, the larger the value of the slope the more 
accurate the determination of the charge radius of the \emph{unstable} 
neutron-deficient nickel isotope needs to be. Indeed, if one is interested in 
the determination of the neutron skin of ${}^{48}$Ca to a precision of 
0.02\,fm\,\cite{CREX:2013}, then one must measure the charge radius 
of ${}^{54}$Ni to better than 0.004\,fm; note that the charge radius of its 
\emph{stable} mirror nucleus ${}^{54}$Fe is already known to 
0.002\,fm\,\cite{Angeli:2013}. On the other hand, for the $A\!=\!50$ 
case the charge radius of ${}^{50}$Ni needs to be determined to ``only" 
0.012\,fm. However, in this case the experimental challenge is formidable, as 
${}^{58}$Ni is the most neutron-deficient isotope with a well measured 
charge radius\,\cite{Angeli:2013}. Yet, we are confident that with the commissioning 
of new and more intense radioactive beam facilities, the experimental community 
will continue to rise up to the challenge. Note that while the regression slopes 
almost double for ${}^{208}$Pb, the aim of the PREX-II experiment is to 
determine the neutron radius of ${}^{208}$Pb to 0.06\,fm. However, we caution 
against a possible model dependence of our results, as the correlation between 
the neutron skin of ${}^{208}$Pb and that of ${}^{48}$Ca does not appear to be 
as strong as suggested here; see for example Fig.\,2(b) of 
Ref.\,\cite{Piekarewicz:2012pp}. 

%%%%%%%%%%%%%%%%%%%%%%%%%%%%%%%%%%%%%%%%%%%%%%%%%%%%%
\begin{figure*}[ht]
\smallskip
 \includegraphics[width=0.65\columnwidth]{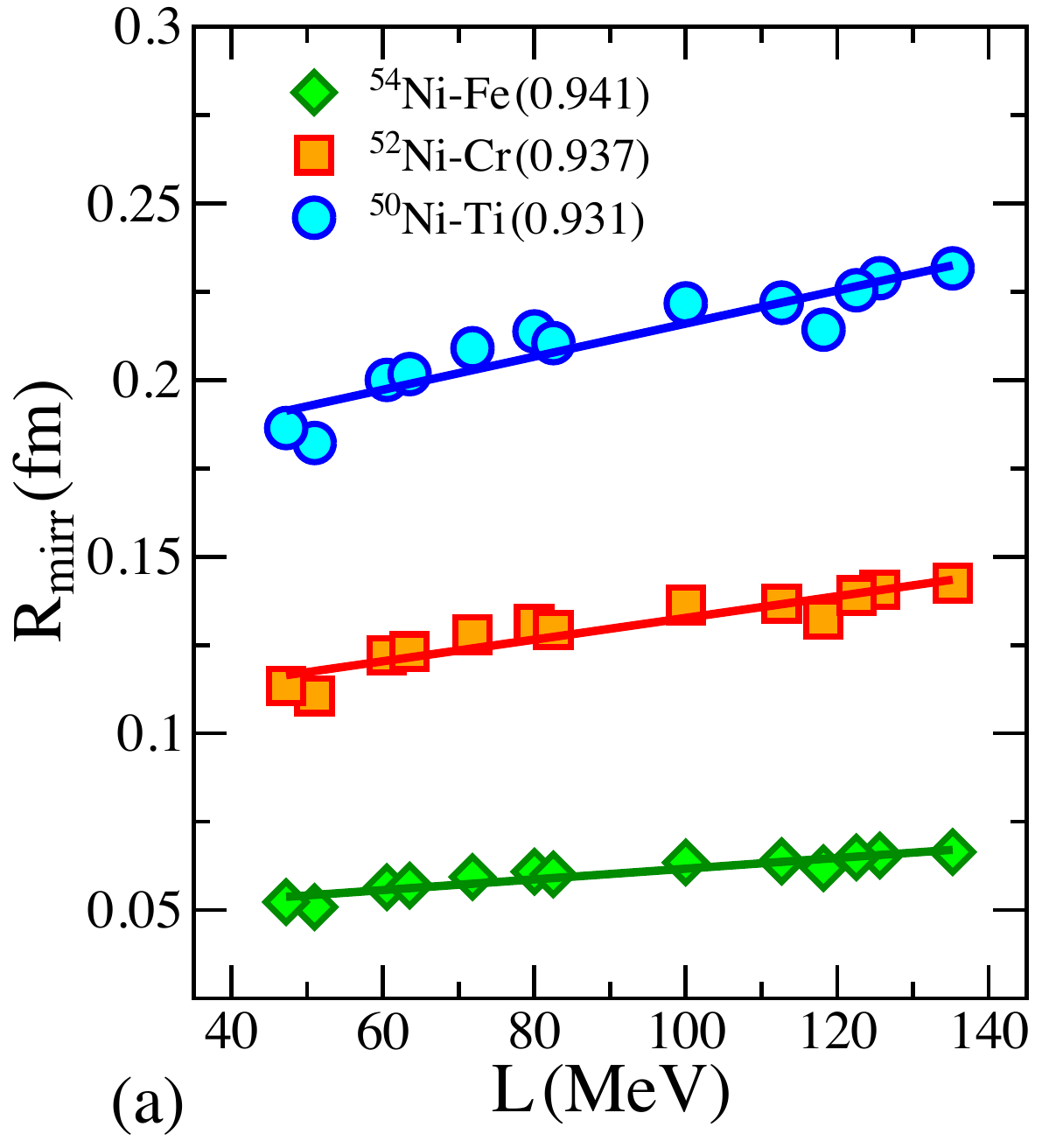}
   \hspace{10pt}
 \includegraphics[width=0.68\columnwidth]{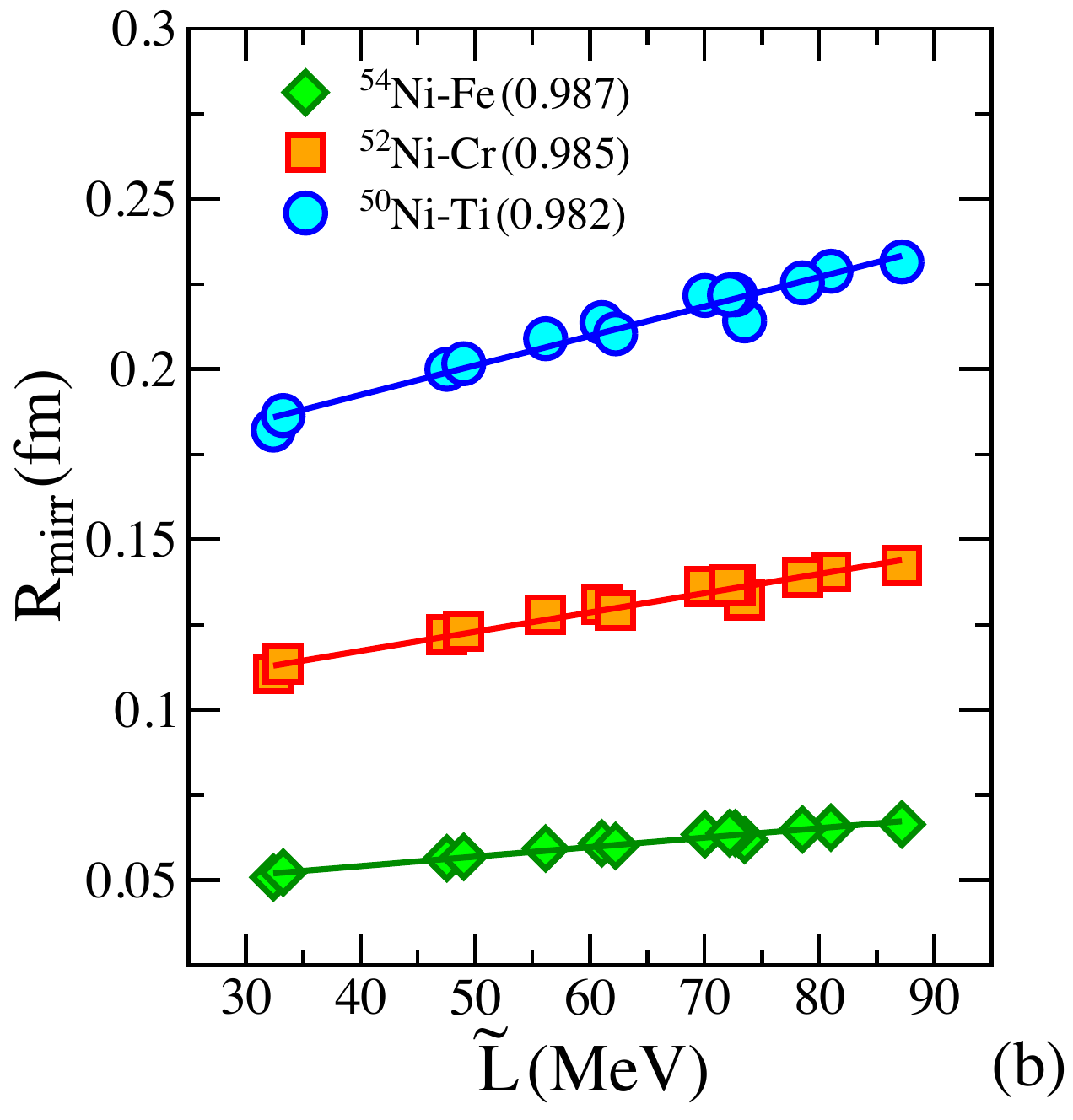}
\caption{(Color online) (a) Difference in proton radii along the 
	      $A\!=\!50, 52, 54$ isobars as a function of the slope 
	      of the symmetry energy at saturation density. Numbers 
	      in parentheses represent the correlation coefficients. 
	      (b) Same but now as a function of the slope of the 
	      symmetry energy at the lower density of $0.10$\,fm.}
\label{Fig2}
\end{figure*}
%%%%%%%%%%%%%%%%%%%%%%%%%%%%%%%%%%%%%%%%%%%%%%%%%%%%%

Having established the existence of a strong correlation between $R_{\rm skin}$ and 
$R_{\rm mirr}$, we now proceed to explore the sensitivity of the latter to the slope of 
the symmetry energy. Recall that the slope of the symmetry energy is defined as
%%%
\begin{equation}
 L = \left(3\rho\frac{\partial\mathcal{S}}
       {\partial\rho}\right)_{\!\!\rho=\rho_{{}_{0}}} \hspace{-10pt},
\label{SlopeL}
\end{equation}
%%%
where $\mathcal{S}(\rho)$ is the symmetry energy, namely, the energy cost 
of turning neutrons into protons (or vice versa) in symmetric nuclear matter. 

In Fig.\,\ref{Fig2}(a) we plot the difference in the proton radii of mirror nuclei 
as a function of L. The observed correlation is as strong as the one between 
the neutron-skin thickness of ${}^{48}$Ca and $L$ (figure not shown). While 
this suggests an efficient tool to constrain a fundamental parameter of the 
equation of state, the robustness of this result should be tested against a
possible model dependency. Clearly, it would be ideal to extend this approach 
to the heavy-mass region where the surface to volume ratio is more favorable, 
as in the case of ${}^{208}$Pb whose neutron skin has been firmly established 
as a proxy for $L$. Unfortunately, exploiting the isovector character of mirror 
nuclei is limited to a fairly narrow region of the nuclear chart. 

Shown in Fig.\,\ref{Fig2}(b) is a similar plot but now against the slope of the 
symmetry energy at the slightly lower density of $\tilde{\rho}_{{}_{0}}\!=\!0.10$\,fm, 
or about 2/3 of the density at saturation. Note that $\widetilde{L}$ is defined exactly 
as in Eq.\,(\ref{SlopeL}) but now evaluated at $\rho\!=\!\tilde\rho_{{}_{0}}$. In all 
three cases the correlation becomes tighter. That a lower density than saturation 
represents a better choice to determine the symmetry energy has been emphasized 
repeatedly; see for example Refs.\,\cite{Farine:1978,Brown:2000,Horowitz:2000xj,
Furnstahl:2001un,Horowitz:2001ya,Ducoin:2011fy,Zhang:2013wna,Brown:2013mga,
Horowitz:2014bja}. Indeed, given that so far the isovector sector is largely informed 
by the binding energy of stable neutron-rich nuclei, the symmetry energy is better 
constrained at a density that results from an average of the nuclear interior and the 
nuclear surface. 

%%%%%%%%%%%%%%%%%%%%%%%%%%%%%%%%%%%%%%%%%%%%%%%%%%%%%
\begin{figure}[ht]
\smallskip
 \includegraphics[width=0.99\columnwidth]{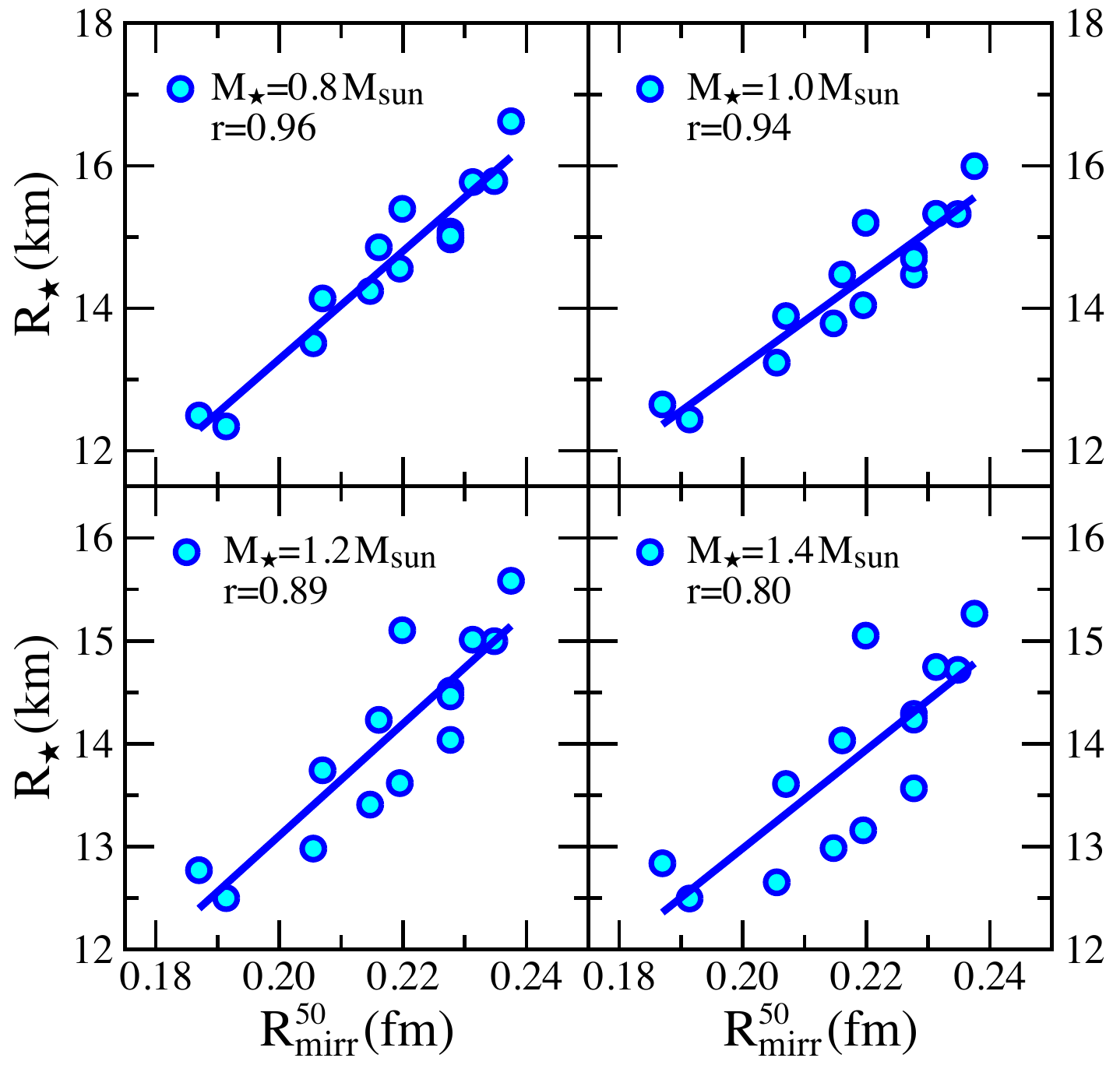}
 \caption{(Color online) Stellar radii for neutron stars with masses
                of $M_{\star}\!=\!0.8,1.0,1.2,1.4\,M_{\odot}$ as a function
                of the difference in proton radii between ${}^{50}$Ni and 
                ${}^{50}$Ti. Here $r$ is the correlation coefficient deduced 
                from a linear regression.}       
\label{Fig3}
\end{figure}
%%%%%%%%%%%%%%%%%%%%%%%%%%%%%%%%%%%%%%%%%%%%%%%%%%%%%

We finish this contribution by exploring a possible connection between 
$R_{\rm mirr}^{50}\!\equiv\!R_{p}({}^{50}{\rm Ni})-R_{p}({}^{50}{\rm Ti})$
and the radius of a neutron star, a stellar property that is known to be 
particularly sensitive to the density dependence of the symmetry 
energy\,\cite{Lattimer:2006xb}. Note that an intriguing correlation 
exists that involves objects that differ in size by 18 orders of magnitude: 
the smaller the neutron-skin thickness of ${}^{208}$Pb the smaller the 
size of the neutron star\,\cite{Horowitz:2001ya}. That is, whether pushing 
against surface tension in ${}^{208}$Pb or against gravity in a ``low-mass" 
neutron star\,\cite{Carriere:2002bx}, it is the pressure of neutron-rich 
matter around saturation density that determines both the thickness of 
the neutron skin and the radius of a neutron star. Given the strong 
correlation between $R_{\rm skin}^{208}$ and $R_{\rm mirr}^{50}$ 
displayed in Fig.\,\ref{Fig1}b, we find natural to explore a possible 
connection between the latter and the stellar radius. Thus, we display
in Fig.\,\ref{Fig3} neutron-star radii as a function of $R_{\rm mirr}^{50}$ 
for neutron stars with masses of $M_{\star}\!=\!0.8,1.0,1.2,1.4\,M_{\odot}$.
We observe a strong correlation---with a correlation coefficient of 
$r=\!0.96$---between $R_{\rm mirr}^{50}$ and the radius of a 
$M_{\star}\!=\!0.8\,M_{\odot}$ neutron star. The correlation is strong 
because for such a relatively light neutron star, the central density 
remains well below twice nuclear-matter saturation density even for 
the softest of the RMF models. However, the correlation weakens with 
increasing stellar mass because the radius becomes sensitive to the 
pressure at densities significantly higher than those probed in the 
laboratory. For example, in the case of a $M_{\star}\!=\!1.4\,M_{\odot}$ 
the correlation weakens to $r=\!0.8$ because now the density in the 
stellar core exceeds three times nuclear matter saturation density. As in 
the case of the neutron-skin thickness of ${}^{208}$Pb, we conclude 
that it may be possible to infer some fundamental properties of low-mass 
neutron stars from the structure of atomic nuclei.

In summary, inspired by the simple and elegant idea presented in
Ref.\,\cite{Brown:2017} that suggests that differences in the charge radii 
of mirror nuclei are correlated to both the neutron-skin thickness of 
neutron-rich nuclei and the slope of the symmetry energy, we have 
investigated the validity of these correlations in the relativistic 
framework. Using a set of accurately calibrated relativistic energy 
density functionals that span a wide range of values for the slope 
of the symmetry energy $L$, we have confirmed the results of 
Ref.\,\cite{Brown:2017}. Moreover, we have extended our results 
to the neutron-star domain and reported a strong correlation between
the difference in the proton radii between ${}^{50}{\rm Ni}$ and
${}^{50}{\rm Ti}$, and the radii of low-mass neutron stars.
Thus, at least within the context of the RMF models employed 
in this work, we have established that the difference in charge 
radii may serve as a credible surrogate for the neutron-skin of
neutron-rich nuclei. Moreover, we concluded that accurate 
measurements of the charge radii of neutron deficient nickel 
isotopes may have important implications for the structure of 
low-mass neutron stars. 

The realization that the neutron-skin thickness of neutron-rich 
nuclei could have such a dramatic impact in areas far beyond
the nuclear-structure domain has created a flurry of activity that 
continues until today. We trust that the ideas introduced in 
Ref.\,\cite{Brown:2017} and expanded in this presentation will 
also stimulate considerable experimental and theoretical activity.
Theoretically, both ab initio models and energy density 
functionals of increasing sophistication are in an excellent
position to predict with quantifiable uncertainties the charge 
distribution of neutron-deficient nuclei. Experimentally, 
enormous technical advances have resulted in pioneering 
measurements of the charge radii of unstable neutron-rich 
isotopes at such facilities as ISOLDE-CERN\,\cite{Ruiz:2016gne} 
and soon at RIKEN-SCRIT\,\cite{Tsukada:2017llu}. We are confident
that these techniques may also be used to measure the charge
radius of the neutron-deficient isotopes discussed in this work.
Moreover, such remarkable level of activity will only increase 
with the commissioning of new radioactive beam facilities 
throughout the world. As we enter a golden era in nuclear 
structure that will see a paradigm shift in fundamental core 
concepts, we are confident that ``unprecedented access to 
a vast new array of nuclei will result in scientific breakthroughs 
and major advances in our understanding of nuclei and their role 
in the cosmos"\,\cite{LongRangePlan}. 

\begin{acknowledgments}
 This material is based upon work supported by the U.S. Department 
 of Energy Office of Science, Office of Nuclear Physics under Award 
 Number DE-FG02-92ER40750.
\end{acknowledgments}

%\bibliography{../ReferencesJP.bib}
\bibliography{MirrirChRadii.bbl}

\begin{thebibliography}{61}
\expandafter\ifx\csname natexlab\endcsname\relax\def\natexlab#1{#1}\fi
\expandafter\ifx\csname bibnamefont\endcsname\relax
  \def\bibnamefont#1{#1}\fi
\expandafter\ifx\csname bibfnamefont\endcsname\relax
  \def\bibfnamefont#1{#1}\fi
\expandafter\ifx\csname citenamefont\endcsname\relax
  \def\citenamefont#1{#1}\fi
\expandafter\ifx\csname url\endcsname\relax
  \def\url#1{\texttt{#1}}\fi
\expandafter\ifx\csname urlprefix\endcsname\relax\def\urlprefix{URL }\fi
\providecommand{\bibinfo}[2]{#2}
\providecommand{\eprint}[2][]{\url{#2}}

\bibitem[{\citenamefont{Brown}(2000)}]{Brown:2000}
\bibinfo{author}{\bibfnamefont{B.~A.} \bibnamefont{Brown}},
  \bibinfo{journal}{Phys. Rev. Lett.} \textbf{\bibinfo{volume}{85}},
  \bibinfo{pages}{5296} (\bibinfo{year}{2000}).

\bibitem[{\citenamefont{Furnstahl}(2002)}]{Furnstahl:2001un}
\bibinfo{author}{\bibfnamefont{R.~J.} \bibnamefont{Furnstahl}},
  \bibinfo{journal}{Nucl. Phys.} \textbf{\bibinfo{volume}{A706}},
  \bibinfo{pages}{85} (\bibinfo{year}{2002}).

\bibitem[{\citenamefont{Centelles et~al.}(2009)\citenamefont{Centelles,
  Roca-Maza, Vi\~nas, and Warda}}]{Centelles:2008vu}
\bibinfo{author}{\bibfnamefont{M.}~\bibnamefont{Centelles}},
  \bibinfo{author}{\bibfnamefont{X.}~\bibnamefont{Roca-Maza}},
  \bibinfo{author}{\bibfnamefont{X.}~\bibnamefont{Vi\~nas}}, \bibnamefont{and}
  \bibinfo{author}{\bibfnamefont{M.}~\bibnamefont{Warda}},
  \bibinfo{journal}{Phys. Rev. Lett.} \textbf{\bibinfo{volume}{102}},
  \bibinfo{pages}{122502} (\bibinfo{year}{2009}).

\bibitem[{\citenamefont{Roca-Maza et~al.}(2011)\citenamefont{Roca-Maza,
  Centelles, Vi\~nas, and Warda}}]{RocaMaza:2011pm}
\bibinfo{author}{\bibfnamefont{X.}~\bibnamefont{Roca-Maza}},
  \bibinfo{author}{\bibfnamefont{M.}~\bibnamefont{Centelles}},
  \bibinfo{author}{\bibfnamefont{X.}~\bibnamefont{Vi\~nas}}, \bibnamefont{and}
  \bibinfo{author}{\bibfnamefont{M.}~\bibnamefont{Warda}},
  \bibinfo{journal}{Phys. Rev. Lett.} \textbf{\bibinfo{volume}{106}},
  \bibinfo{pages}{252501} (\bibinfo{year}{2011}).

\bibitem[{\citenamefont{Donnelly et~al.}(1989)\citenamefont{Donnelly, Dubach,
  and Sick}}]{Donnelly:1989qs}
\bibinfo{author}{\bibfnamefont{T.}~\bibnamefont{Donnelly}},
  \bibinfo{author}{\bibfnamefont{J.}~\bibnamefont{Dubach}}, \bibnamefont{and}
  \bibinfo{author}{\bibfnamefont{I.}~\bibnamefont{Sick}},
  \bibinfo{journal}{Nucl.Phys.} \textbf{\bibinfo{volume}{A503}},
  \bibinfo{pages}{589} (\bibinfo{year}{1989}).

\bibitem[{\citenamefont{Abrahamyan et~al.}(2012)\citenamefont{Abrahamyan,
  Ahmed, Albataineh, Aniol, Armstrong et~al.}}]{Abrahamyan:2012gp}
\bibinfo{author}{\bibfnamefont{S.}~\bibnamefont{Abrahamyan}},
  \bibinfo{author}{\bibfnamefont{Z.}~\bibnamefont{Ahmed}},
  \bibinfo{author}{\bibfnamefont{H.}~\bibnamefont{Albataineh}},
  \bibinfo{author}{\bibfnamefont{K.}~\bibnamefont{Aniol}},
  \bibinfo{author}{\bibfnamefont{D.~S.} \bibnamefont{Armstrong}},
  \bibnamefont{et~al.}, \bibinfo{journal}{Phys. Rev. Lett.}
  \textbf{\bibinfo{volume}{108}}, \bibinfo{pages}{112502}
  (\bibinfo{year}{2012}).

\bibitem[{\citenamefont{Horowitz et~al.}(2012)\citenamefont{Horowitz, Ahmed,
  Jen, Rakhman, Souder et~al.}}]{Horowitz:2012tj}
\bibinfo{author}{\bibfnamefont{C.~J.} \bibnamefont{Horowitz}},
  \bibinfo{author}{\bibfnamefont{Z.}~\bibnamefont{Ahmed}},
  \bibinfo{author}{\bibfnamefont{C.~M.} \bibnamefont{Jen}},
  \bibinfo{author}{\bibfnamefont{A.}~\bibnamefont{Rakhman}},
  \bibinfo{author}{\bibfnamefont{P.~A.} \bibnamefont{Souder}},
  \bibnamefont{et~al.}, \bibinfo{journal}{Phys. Rev.}
  \textbf{\bibinfo{volume}{C85}}, \bibinfo{pages}{032501}
  (\bibinfo{year}{2012}).

\bibitem[{\citenamefont{Mammei et~al.}(2013)}]{CREX:2013}
\bibinfo{author}{\bibfnamefont{J.}~\bibnamefont{Mammei}} \bibnamefont{et~al.},
  \emph{\bibinfo{title}{\uppercase{CREX}: Parity-violating measurement of the
  weak-charge distribution of $\textsuperscript{48}$\uppercase{C}a to 0.02\,fm
  accuracy}} (\bibinfo{year}{2013}),
  \urlprefix\url{http://hallaweb.jlab.org/parity/prex/c-rex/c-rex.pdf}.

\bibitem[{\citenamefont{Balantekin et~al.}(2014)}]{Balantekin:2014opa}
\bibinfo{author}{\bibfnamefont{A.~B.} \bibnamefont{Balantekin}}
  \bibnamefont{et~al.}, \bibinfo{journal}{Mod. Phys. Lett.}
  \textbf{\bibinfo{volume}{A29}}, \bibinfo{pages}{1430010}
  (\bibinfo{year}{2014}).

\bibitem[{Lon(2015)}]{LongRangePlan}
\emph{\bibinfo{title}{Reaching for the Horizon; The 2015 Long Range Plan for
  Nuclear Science}} (\bibinfo{year}{2015}).

\bibitem[{\citenamefont{Tsang et~al.}(2004)}]{Tsang:2004zz}
\bibinfo{author}{\bibfnamefont{M.~B.} \bibnamefont{Tsang}}
  \bibnamefont{et~al.}, \bibinfo{journal}{Phys. Rev. Lett.}
  \textbf{\bibinfo{volume}{92}}, \bibinfo{pages}{062701}
  (\bibinfo{year}{2004}).

\bibitem[{\citenamefont{Chen et~al.}(2005)\citenamefont{Chen, Ko, and
  Li}}]{Chen:2004si}
\bibinfo{author}{\bibfnamefont{L.-W.} \bibnamefont{Chen}},
  \bibinfo{author}{\bibfnamefont{C.~M.} \bibnamefont{Ko}}, \bibnamefont{and}
  \bibinfo{author}{\bibfnamefont{B.-A.} \bibnamefont{Li}},
  \bibinfo{journal}{Phys. Rev. Lett.} \textbf{\bibinfo{volume}{94}},
  \bibinfo{pages}{032701} (\bibinfo{year}{2005}).

\bibitem[{\citenamefont{Steiner and Li}(2005)}]{Steiner:2005rd}
\bibinfo{author}{\bibfnamefont{A.~W.} \bibnamefont{Steiner}} \bibnamefont{and}
  \bibinfo{author}{\bibfnamefont{B.-A.} \bibnamefont{Li}},
  \bibinfo{journal}{Phys. Rev.} \textbf{\bibinfo{volume}{C72}},
  \bibinfo{pages}{041601} (\bibinfo{year}{2005}).

\bibitem[{\citenamefont{Shetty et~al.}(2007)\citenamefont{Shetty, Yennello, and
  Souliotis}}]{Shetty:2007zg}
\bibinfo{author}{\bibfnamefont{D.~V.} \bibnamefont{Shetty}},
  \bibinfo{author}{\bibfnamefont{S.~J.} \bibnamefont{Yennello}},
  \bibnamefont{and} \bibinfo{author}{\bibfnamefont{G.~A.}
  \bibnamefont{Souliotis}}, \bibinfo{journal}{Phys. Rev.}
  \textbf{\bibinfo{volume}{C76}}, \bibinfo{pages}{024606}
  (\bibinfo{year}{2007}).

\bibitem[{\citenamefont{Tsang et~al.}(2009)}]{Tsang:2008fd}
\bibinfo{author}{\bibfnamefont{M.~B.} \bibnamefont{Tsang}}
  \bibnamefont{et~al.}, \bibinfo{journal}{Phys. Rev. Lett.}
  \textbf{\bibinfo{volume}{102}}, \bibinfo{pages}{122701}
  (\bibinfo{year}{2009}).

\bibitem[{\citenamefont{Li et~al.}(2008)\citenamefont{Li, Chen, and
  Ko}}]{Li:2008gp}
\bibinfo{author}{\bibfnamefont{B.-A.} \bibnamefont{Li}},
  \bibinfo{author}{\bibfnamefont{L.-W.} \bibnamefont{Chen}}, \bibnamefont{and}
  \bibinfo{author}{\bibfnamefont{C.~M.} \bibnamefont{Ko}},
  \bibinfo{journal}{Phys. Rept.} \textbf{\bibinfo{volume}{464}},
  \bibinfo{pages}{113} (\bibinfo{year}{2008}).

\bibitem[{\citenamefont{Tsang et~al.}(2012)\citenamefont{Tsang, Stone, Camera,
  Danielewicz, Gandolfi et~al.}}]{Tsang:2012se}
\bibinfo{author}{\bibfnamefont{M.}~\bibnamefont{Tsang}},
  \bibinfo{author}{\bibfnamefont{J.}~\bibnamefont{Stone}},
  \bibinfo{author}{\bibfnamefont{F.}~\bibnamefont{Camera}},
  \bibinfo{author}{\bibfnamefont{P.}~\bibnamefont{Danielewicz}},
  \bibinfo{author}{\bibfnamefont{S.}~\bibnamefont{Gandolfi}},
  \bibnamefont{et~al.}, \bibinfo{journal}{Phys.Rev.}
  \textbf{\bibinfo{volume}{C86}}, \bibinfo{pages}{015803}
  (\bibinfo{year}{2012}).

\bibitem[{\citenamefont{Horowitz et~al.}(2014)\citenamefont{Horowitz, Brown,
  Kim, Lynch, Michaels et~al.}}]{Horowitz:2014bja}
\bibinfo{author}{\bibfnamefont{C.~J.} \bibnamefont{Horowitz}},
  \bibinfo{author}{\bibfnamefont{E.~F.} \bibnamefont{Brown}},
  \bibinfo{author}{\bibfnamefont{Y.}~\bibnamefont{Kim}},
  \bibinfo{author}{\bibfnamefont{W.~G.} \bibnamefont{Lynch}},
  \bibinfo{author}{\bibfnamefont{R.}~\bibnamefont{Michaels}},
  \bibnamefont{et~al.}, \bibinfo{journal}{J. Phys.}
  \textbf{\bibinfo{volume}{G41}}, \bibinfo{pages}{093001}
  (\bibinfo{year}{2014}).

\bibitem[{\citenamefont{Horowitz and
  Piekarewicz}(2001{\natexlab{a}})}]{Horowitz:2000xj}
\bibinfo{author}{\bibfnamefont{C.~J.} \bibnamefont{Horowitz}} \bibnamefont{and}
  \bibinfo{author}{\bibfnamefont{J.}~\bibnamefont{Piekarewicz}},
  \bibinfo{journal}{Phys. Rev. Lett.} \textbf{\bibinfo{volume}{86}},
  \bibinfo{pages}{5647} (\bibinfo{year}{2001}{\natexlab{a}}).

\bibitem[{\citenamefont{Horowitz and
  Piekarewicz}(2001{\natexlab{b}})}]{Horowitz:2001ya}
\bibinfo{author}{\bibfnamefont{C.~J.} \bibnamefont{Horowitz}} \bibnamefont{and}
  \bibinfo{author}{\bibfnamefont{J.}~\bibnamefont{Piekarewicz}},
  \bibinfo{journal}{Phys. Rev.} \textbf{\bibinfo{volume}{C64}},
  \bibinfo{pages}{062802} (\bibinfo{year}{2001}{\natexlab{b}}).

\bibitem[{\citenamefont{Carriere et~al.}(2003)\citenamefont{Carriere, Horowitz,
  and Piekarewicz}}]{Carriere:2002bx}
\bibinfo{author}{\bibfnamefont{J.}~\bibnamefont{Carriere}},
  \bibinfo{author}{\bibfnamefont{C.~J.} \bibnamefont{Horowitz}},
  \bibnamefont{and}
  \bibinfo{author}{\bibfnamefont{J.}~\bibnamefont{Piekarewicz}},
  \bibinfo{journal}{Astrophys. J.} \textbf{\bibinfo{volume}{593}},
  \bibinfo{pages}{463} (\bibinfo{year}{2003}).

\bibitem[{\citenamefont{Steiner et~al.}(2005)\citenamefont{Steiner, Prakash,
  Lattimer, and Ellis}}]{Steiner:2004fi}
\bibinfo{author}{\bibfnamefont{A.~W.} \bibnamefont{Steiner}},
  \bibinfo{author}{\bibfnamefont{M.}~\bibnamefont{Prakash}},
  \bibinfo{author}{\bibfnamefont{J.~M.} \bibnamefont{Lattimer}},
  \bibnamefont{and} \bibinfo{author}{\bibfnamefont{P.~J.} \bibnamefont{Ellis}},
  \bibinfo{journal}{Phys. Rept.} \textbf{\bibinfo{volume}{411}},
  \bibinfo{pages}{325} (\bibinfo{year}{2005}).

\bibitem[{\citenamefont{Li and Steiner}(2006)}]{Li:2005sr}
\bibinfo{author}{\bibfnamefont{B.-A.} \bibnamefont{Li}} \bibnamefont{and}
  \bibinfo{author}{\bibfnamefont{A.~W.} \bibnamefont{Steiner}},
  \bibinfo{journal}{Phys. Lett.} \textbf{\bibinfo{volume}{B642}},
  \bibinfo{pages}{436} (\bibinfo{year}{2006}).

\bibitem[{\citenamefont{Erler et~al.}(2013)\citenamefont{Erler, Horowitz,
  Nazarewicz, Rafalski, and Reinhard}}]{Erler:2012qd}
\bibinfo{author}{\bibfnamefont{J.}~\bibnamefont{Erler}},
  \bibinfo{author}{\bibfnamefont{C.~J.} \bibnamefont{Horowitz}},
  \bibinfo{author}{\bibfnamefont{W.}~\bibnamefont{Nazarewicz}},
  \bibinfo{author}{\bibfnamefont{M.}~\bibnamefont{Rafalski}}, \bibnamefont{and}
  \bibinfo{author}{\bibfnamefont{P.-G.} \bibnamefont{Reinhard}},
  \bibinfo{journal}{Phys. Rev.} \textbf{\bibinfo{volume}{C87}},
  \bibinfo{pages}{044320} (\bibinfo{year}{2013}).

\bibitem[{\citenamefont{Chen and Piekarewicz}(2014)}]{Chen:2014sca}
\bibinfo{author}{\bibfnamefont{W.-C.} \bibnamefont{Chen}} \bibnamefont{and}
  \bibinfo{author}{\bibfnamefont{J.}~\bibnamefont{Piekarewicz}},
  \bibinfo{journal}{Phys. Rev.} \textbf{\bibinfo{volume}{C90}},
  \bibinfo{pages}{044305} (\bibinfo{year}{2014}).

\bibitem[{\citenamefont{Reinhard and Nazarewicz}(2010)}]{Reinhard:2010wz}
\bibinfo{author}{\bibfnamefont{P.-G.} \bibnamefont{Reinhard}} \bibnamefont{and}
  \bibinfo{author}{\bibfnamefont{W.}~\bibnamefont{Nazarewicz}},
  \bibinfo{journal}{Phys. Rev.} \textbf{\bibinfo{volume}{C81}},
  \bibinfo{pages}{051303} (\bibinfo{year}{2010}).

\bibitem[{\citenamefont{Harakeh and van~der Woude}(2001)}]{Harakeh:2001}
\bibinfo{author}{\bibfnamefont{M.~N.} \bibnamefont{Harakeh}} \bibnamefont{and}
  \bibinfo{author}{\bibfnamefont{A.}~\bibnamefont{van~der Woude}},
  \emph{\bibinfo{title}{Giant Resonances-Fundamental High-frequency Modes of
  Nuclear Excitation}} (\bibinfo{publisher}{Clarendon, Oxford},
  \bibinfo{year}{2001}).

\bibitem[{\citenamefont{Piekarewicz}(2011)}]{Piekarewicz:2010fa}
\bibinfo{author}{\bibfnamefont{J.}~\bibnamefont{Piekarewicz}},
  \bibinfo{journal}{Phys. Rev.} \textbf{\bibinfo{volume}{C83}},
  \bibinfo{pages}{034319} (\bibinfo{year}{2011}).

\bibitem[{\citenamefont{Piekarewicz et~al.}(2012)\citenamefont{Piekarewicz,
  Agrawal, Col\`o, Nazarewicz, Paar et~al.}}]{Piekarewicz:2012pp}
\bibinfo{author}{\bibfnamefont{J.}~\bibnamefont{Piekarewicz}},
  \bibinfo{author}{\bibfnamefont{B.}~\bibnamefont{Agrawal}},
  \bibinfo{author}{\bibfnamefont{G.}~\bibnamefont{Col\`o}},
  \bibinfo{author}{\bibfnamefont{W.}~\bibnamefont{Nazarewicz}},
  \bibinfo{author}{\bibfnamefont{N.}~\bibnamefont{Paar}}, \bibnamefont{et~al.},
  \bibinfo{journal}{Phys. Rev.} \textbf{\bibinfo{volume}{C85}},
  \bibinfo{pages}{041302(R)} (\bibinfo{year}{2012}).

\bibitem[{\citenamefont{Reinhard and Nazarewicz}(2013)}]{Reinhard:2012vw}
\bibinfo{author}{\bibfnamefont{P.}~\bibnamefont{Reinhard}} \bibnamefont{and}
  \bibinfo{author}{\bibfnamefont{W.}~\bibnamefont{Nazarewicz}},
  \bibinfo{journal}{Phys. Rev.} \textbf{\bibinfo{volume}{C87}},
  \bibinfo{pages}{014324} (\bibinfo{year}{2013}).

\bibitem[{\citenamefont{Roca-Maza et~al.}(2013)\citenamefont{Roca-Maza,
  Centelles, Vi\~nas, Brenna, Col\`o et~al.}}]{Roca-Maza:2013mla}
\bibinfo{author}{\bibfnamefont{X.}~\bibnamefont{Roca-Maza}},
  \bibinfo{author}{\bibfnamefont{M.}~\bibnamefont{Centelles}},
  \bibinfo{author}{\bibfnamefont{X.}~\bibnamefont{Vi\~nas}},
  \bibinfo{author}{\bibfnamefont{M.}~\bibnamefont{Brenna}},
  \bibinfo{author}{\bibfnamefont{G.}~\bibnamefont{Col\`o}},
  \bibnamefont{et~al.}, \bibinfo{journal}{Phys. Rev.}
  \textbf{\bibinfo{volume}{C88}}, \bibinfo{pages}{024316}
  (\bibinfo{year}{2013}).

\bibitem[{\citenamefont{Roca-Maza et~al.}(2015)\citenamefont{Roca-Maza,
  Vi\~nas, Centelles, Agrawal, Col\`o, Paar, Piekarewicz, and
  Vretenar}}]{Roca-Maza:2015eza}
\bibinfo{author}{\bibfnamefont{X.}~\bibnamefont{Roca-Maza}},
  \bibinfo{author}{\bibfnamefont{X.}~\bibnamefont{Vi\~nas}},
  \bibinfo{author}{\bibfnamefont{M.}~\bibnamefont{Centelles}},
  \bibinfo{author}{\bibfnamefont{B.~K.} \bibnamefont{Agrawal}},
  \bibinfo{author}{\bibfnamefont{G.}~\bibnamefont{Col\`o}},
  \bibinfo{author}{\bibfnamefont{N.}~\bibnamefont{Paar}},
  \bibinfo{author}{\bibfnamefont{J.}~\bibnamefont{Piekarewicz}},
  \bibnamefont{and} \bibinfo{author}{\bibfnamefont{D.}~\bibnamefont{Vretenar}},
  \bibinfo{journal}{Phys. Rev.} \textbf{\bibinfo{volume}{C92}},
  \bibinfo{pages}{064304} (\bibinfo{year}{2015}).

\bibitem[{\citenamefont{Hagen et~al.}(2015)}]{Hagen:2015yea}
\bibinfo{author}{\bibfnamefont{G.}~\bibnamefont{Hagen}} \bibnamefont{et~al.},
  \bibinfo{journal}{Nature Phys.}  (\bibinfo{year}{2015}).

\bibitem[{\citenamefont{Piekarewicz}(2006)}]{Piekarewicz:2006ip}
\bibinfo{author}{\bibfnamefont{J.}~\bibnamefont{Piekarewicz}},
  \bibinfo{journal}{Phys. Rev.} \textbf{\bibinfo{volume}{C73}},
  \bibinfo{pages}{044325} (\bibinfo{year}{2006}).

\bibitem[{\citenamefont{Tamii et~al.}(2011)}]{Tamii:2011pv}
\bibinfo{author}{\bibfnamefont{A.}~\bibnamefont{Tamii}} \bibnamefont{et~al.},
  \bibinfo{journal}{Phys. Rev. Lett.} \textbf{\bibinfo{volume}{107}},
  \bibinfo{pages}{062502} (\bibinfo{year}{2011}).

\bibitem[{\citenamefont{Poltoratska et~al.}(2012)\citenamefont{Poltoratska, von
  Neumann-Cosel, Tamii, Adachi, Bertulani et~al.}}]{Poltoratska:2012nf}
\bibinfo{author}{\bibfnamefont{I.}~\bibnamefont{Poltoratska}},
  \bibinfo{author}{\bibfnamefont{P.}~\bibnamefont{von Neumann-Cosel}},
  \bibinfo{author}{\bibfnamefont{A.}~\bibnamefont{Tamii}},
  \bibinfo{author}{\bibfnamefont{T.}~\bibnamefont{Adachi}},
  \bibinfo{author}{\bibfnamefont{C.}~\bibnamefont{Bertulani}},
  \bibnamefont{et~al.}, \bibinfo{journal}{Phys. Rev.}
  \textbf{\bibinfo{volume}{C85}}, \bibinfo{pages}{041304}
  (\bibinfo{year}{2012}).

\bibitem[{\citenamefont{Tamii et~al.}(2014)\citenamefont{Tamii, von
  Neumann-Cosel, and Poltoratska}}]{Tamii:2013cna}
\bibinfo{author}{\bibfnamefont{A.}~\bibnamefont{Tamii}},
  \bibinfo{author}{\bibfnamefont{P.}~\bibnamefont{von Neumann-Cosel}},
  \bibnamefont{and}
  \bibinfo{author}{\bibfnamefont{I.}~\bibnamefont{Poltoratska}},
  \bibinfo{journal}{Eur. Phys. J.} \textbf{\bibinfo{volume}{A50}},
  \bibinfo{pages}{28} (\bibinfo{year}{2014}).

\bibitem[{\citenamefont{Savran et~al.}(2013)\citenamefont{Savran, Aumann, and
  Zilges}}]{Savran:2013bha}
\bibinfo{author}{\bibfnamefont{D.}~\bibnamefont{Savran}},
  \bibinfo{author}{\bibfnamefont{T.}~\bibnamefont{Aumann}}, \bibnamefont{and}
  \bibinfo{author}{\bibfnamefont{A.}~\bibnamefont{Zilges}},
  \bibinfo{journal}{Prog. Part. Nucl. Phys.} \textbf{\bibinfo{volume}{70}},
  \bibinfo{pages}{210} (\bibinfo{year}{2013}).

\bibitem[{\citenamefont{Hashimoto et~al.}(2015)}]{Hashimoto:2015ema}
\bibinfo{author}{\bibfnamefont{T.}~\bibnamefont{Hashimoto}}
  \bibnamefont{et~al.}, \bibinfo{journal}{Phys. Rev.}
  \textbf{\bibinfo{volume}{C92}}, \bibinfo{pages}{031305}
  (\bibinfo{year}{2015}).

\bibitem[{\citenamefont{Rossi et~al.}(2013)\citenamefont{Rossi, Adrich, Aksouh,
  Alvarez-Pol, Aumann et~al.}}]{Rossi:2013xha}
\bibinfo{author}{\bibfnamefont{D.}~\bibnamefont{Rossi}},
  \bibinfo{author}{\bibfnamefont{P.}~\bibnamefont{Adrich}},
  \bibinfo{author}{\bibfnamefont{F.}~\bibnamefont{Aksouh}},
  \bibinfo{author}{\bibfnamefont{H.}~\bibnamefont{Alvarez-Pol}},
  \bibinfo{author}{\bibfnamefont{T.}~\bibnamefont{Aumann}},
  \bibnamefont{et~al.}, \bibinfo{journal}{Phys. Rev. Lett.}
  \textbf{\bibinfo{volume}{111}}, \bibinfo{pages}{242503}
  (\bibinfo{year}{2013}).

\bibitem[{\citenamefont{Birkhan et~al.}(2017)}]{Birkhan:2016qkr}
\bibinfo{author}{\bibfnamefont{J.}~\bibnamefont{Birkhan}} \bibnamefont{et~al.},
  \bibinfo{journal}{Phys. Rev. Lett.} \textbf{\bibinfo{volume}{118}},
  \bibinfo{pages}{252501} (\bibinfo{year}{2017}).

\bibitem[{\citenamefont{Tonchev et~al.}(2017)}]{Tonchev:2017ily}
\bibinfo{author}{\bibfnamefont{A.~P.} \bibnamefont{Tonchev}}
  \bibnamefont{et~al.}, \bibinfo{journal}{Phys. Lett.}
  \textbf{\bibinfo{volume}{B773}}, \bibinfo{pages}{20} (\bibinfo{year}{2017}).

\bibitem[{\citenamefont{Brown}(2017)}]{Brown:2017}
\bibinfo{author}{\bibfnamefont{B.~A.} \bibnamefont{Brown}},
  \bibinfo{journal}{Phys. Rev. Lett.} \textbf{\bibinfo{volume}{119}},
  \bibinfo{pages}{122502} (\bibinfo{year}{2017}).

\bibitem[{\citenamefont{Walecka}(1974)}]{Walecka:1974qa}
\bibinfo{author}{\bibfnamefont{J.~D.} \bibnamefont{Walecka}},
  \bibinfo{journal}{Annals Phys.} \textbf{\bibinfo{volume}{83}},
  \bibinfo{pages}{491} (\bibinfo{year}{1974}).

\bibitem[{\citenamefont{Serot and Walecka}(1986)}]{Serot:1984ey}
\bibinfo{author}{\bibfnamefont{B.~D.} \bibnamefont{Serot}} \bibnamefont{and}
  \bibinfo{author}{\bibfnamefont{J.~D.} \bibnamefont{Walecka}},
  \bibinfo{journal}{Adv. Nucl. Phys.} \textbf{\bibinfo{volume}{16}},
  \bibinfo{pages}{1} (\bibinfo{year}{1986}).

\bibitem[{\citenamefont{Boguta and Bodmer}(1977)}]{Boguta:1977xi}
\bibinfo{author}{\bibfnamefont{J.}~\bibnamefont{Boguta}} \bibnamefont{and}
  \bibinfo{author}{\bibfnamefont{A.~R.} \bibnamefont{Bodmer}},
  \bibinfo{journal}{Nucl. Phys.} \textbf{\bibinfo{volume}{A292}},
  \bibinfo{pages}{413} (\bibinfo{year}{1977}).

\bibitem[{\citenamefont{Mueller and Serot}(1996)}]{Mueller:1996pm}
\bibinfo{author}{\bibfnamefont{H.}~\bibnamefont{Mueller}} \bibnamefont{and}
  \bibinfo{author}{\bibfnamefont{B.~D.} \bibnamefont{Serot}},
  \bibinfo{journal}{Nucl. Phys.} \textbf{\bibinfo{volume}{A606}},
  \bibinfo{pages}{508} (\bibinfo{year}{1996}).

\bibitem[{\citenamefont{Lalazissis et~al.}(1997)\citenamefont{Lalazissis,
  Konig, and Ring}}]{Lalazissis:1996rd}
\bibinfo{author}{\bibfnamefont{G.~A.} \bibnamefont{Lalazissis}},
  \bibinfo{author}{\bibfnamefont{J.}~\bibnamefont{Konig}}, \bibnamefont{and}
  \bibinfo{author}{\bibfnamefont{P.}~\bibnamefont{Ring}},
  \bibinfo{journal}{Phys. Rev.} \textbf{\bibinfo{volume}{C55}},
  \bibinfo{pages}{540} (\bibinfo{year}{1997}).

\bibitem[{\citenamefont{Lalazissis et~al.}(1999)\citenamefont{Lalazissis,
  Raman, and Ring}}]{Lalazissis:1999}
\bibinfo{author}{\bibfnamefont{G.~A.} \bibnamefont{Lalazissis}},
  \bibinfo{author}{\bibfnamefont{S.}~\bibnamefont{Raman}}, \bibnamefont{and}
  \bibinfo{author}{\bibfnamefont{P.}~\bibnamefont{Ring}}, \bibinfo{journal}{At.
  Data Nucl. Data Tables} \textbf{\bibinfo{volume}{71}}, \bibinfo{pages}{1}
  (\bibinfo{year}{1999}).

\bibitem[{\citenamefont{Todd-Rutel and Piekarewicz}(2005)}]{Todd-Rutel:2005fa}
\bibinfo{author}{\bibfnamefont{B.~G.} \bibnamefont{Todd-Rutel}}
  \bibnamefont{and}
  \bibinfo{author}{\bibfnamefont{J.}~\bibnamefont{Piekarewicz}},
  \bibinfo{journal}{Phys. Rev. Lett} \textbf{\bibinfo{volume}{95}},
  \bibinfo{pages}{122501} (\bibinfo{year}{2005}).

\bibitem[{\citenamefont{Fattoyev et~al.}(2010)\citenamefont{Fattoyev, Horowitz,
  Piekarewicz, and Shen}}]{Fattoyev:2010mx}
\bibinfo{author}{\bibfnamefont{F.~J.} \bibnamefont{Fattoyev}},
  \bibinfo{author}{\bibfnamefont{C.~J.} \bibnamefont{Horowitz}},
  \bibinfo{author}{\bibfnamefont{J.}~\bibnamefont{Piekarewicz}},
  \bibnamefont{and} \bibinfo{author}{\bibfnamefont{G.}~\bibnamefont{Shen}},
  \bibinfo{journal}{Phys. Rev.} \textbf{\bibinfo{volume}{C82}},
  \bibinfo{pages}{055803} (\bibinfo{year}{2010}).

\bibitem[{\citenamefont{Fattoyev and Piekarewicz}(2013)}]{Fattoyev:2013yaa}
\bibinfo{author}{\bibfnamefont{F.}~\bibnamefont{Fattoyev}} \bibnamefont{and}
  \bibinfo{author}{\bibfnamefont{J.}~\bibnamefont{Piekarewicz}},
  \bibinfo{journal}{Phys. Rev. Lett.} \textbf{\bibinfo{volume}{111}},
  \bibinfo{pages}{162501} (\bibinfo{year}{2013}).

\bibitem[{\citenamefont{Chen and Piekarewicz}(2015)}]{Chen:2014mza}
\bibinfo{author}{\bibfnamefont{W.-C.} \bibnamefont{Chen}} \bibnamefont{and}
  \bibinfo{author}{\bibfnamefont{J.}~\bibnamefont{Piekarewicz}},
  \bibinfo{journal}{Phys. Lett.} \textbf{\bibinfo{volume}{B748}},
  \bibinfo{pages}{284} (\bibinfo{year}{2015}).

\bibitem[{\citenamefont{Angeli and Marinova}(2013)}]{Angeli:2013}
\bibinfo{author}{\bibfnamefont{I.}~\bibnamefont{Angeli}} \bibnamefont{and}
  \bibinfo{author}{\bibfnamefont{K.}~\bibnamefont{Marinova}},
  \bibinfo{journal}{At. Data Nucl. Data Tables} \textbf{\bibinfo{volume}{99}},
  \bibinfo{pages}{69 } (\bibinfo{year}{2013}).

\bibitem[{\citenamefont{Farine et~al.}(1978)\citenamefont{Farine, Pearson, and
  Rouben}}]{Farine:1978}
\bibinfo{author}{\bibfnamefont{M.}~\bibnamefont{Farine}},
  \bibinfo{author}{\bibfnamefont{J.}~\bibnamefont{Pearson}}, \bibnamefont{and}
  \bibinfo{author}{\bibfnamefont{B.}~\bibnamefont{Rouben}},
  \bibinfo{journal}{Nucl. Phys. A} \textbf{\bibinfo{volume}{304}},
  \bibinfo{pages}{317 } (\bibinfo{year}{1978}).

\bibitem[{\citenamefont{Ducoin et~al.}(2011)\citenamefont{Ducoin, Margueron,
  Providencia, and Vidana}}]{Ducoin:2011fy}
\bibinfo{author}{\bibfnamefont{C.}~\bibnamefont{Ducoin}},
  \bibinfo{author}{\bibfnamefont{J.}~\bibnamefont{Margueron}},
  \bibinfo{author}{\bibfnamefont{C.}~\bibnamefont{Providencia}},
  \bibnamefont{and} \bibinfo{author}{\bibfnamefont{I.}~\bibnamefont{Vidana}},
  \bibinfo{journal}{Phys.Rev.} \textbf{\bibinfo{volume}{C83}},
  \bibinfo{pages}{045810} (\bibinfo{year}{2011}).

\bibitem[{\citenamefont{Zhang and Chen}(2013)}]{Zhang:2013wna}
\bibinfo{author}{\bibfnamefont{Z.}~\bibnamefont{Zhang}} \bibnamefont{and}
  \bibinfo{author}{\bibfnamefont{L.-W.} \bibnamefont{Chen}},
  \bibinfo{journal}{Phys. Lett.} \textbf{\bibinfo{volume}{B726}},
  \bibinfo{pages}{234} (\bibinfo{year}{2013}).

\bibitem[{\citenamefont{Brown}(2013)}]{Brown:2013mga}
\bibinfo{author}{\bibfnamefont{B.~A.} \bibnamefont{Brown}},
  \bibinfo{journal}{Phys. Rev. Lett.} \textbf{\bibinfo{volume}{111}},
  \bibinfo{pages}{232502} (\bibinfo{year}{2013}).

\bibitem[{\citenamefont{Lattimer and Prakash}(2007)}]{Lattimer:2006xb}
\bibinfo{author}{\bibfnamefont{J.~M.} \bibnamefont{Lattimer}} \bibnamefont{and}
  \bibinfo{author}{\bibfnamefont{M.}~\bibnamefont{Prakash}},
  \bibinfo{journal}{Phys. Rept.} \textbf{\bibinfo{volume}{442}},
  \bibinfo{pages}{109} (\bibinfo{year}{2007}).

\bibitem[{\citenamefont{Garcia~Ruiz et~al.}(2016)}]{Ruiz:2016gne}
\bibinfo{author}{\bibfnamefont{R.~F.} \bibnamefont{Garcia~Ruiz}}
  \bibnamefont{et~al.}, \bibinfo{journal}{Nature Phys.}
  \textbf{\bibinfo{volume}{12}}, \bibinfo{pages}{594} (\bibinfo{year}{2016}).

\bibitem[{\citenamefont{Tsukada et~al.}(2017)}]{Tsukada:2017llu}
\bibinfo{author}{\bibfnamefont{K.}~\bibnamefont{Tsukada}} \bibnamefont{et~al.},
  \bibinfo{journal}{Phys. Rev. Lett.} \textbf{\bibinfo{volume}{118}},
  \bibinfo{pages}{262501} (\bibinfo{year}{2017}).

\end{thebibliography}
%%%%%%%%%%%%%%%%%%%%%%%%%%%%%%%%%%%%%%%%%%%%%%%%%%%%%%%%%%%%%%%%%
\vfill\eject
\end{document}